\documentclass[twocolumn,aps,prl,showpacs, graphicx]{revtex4-1}

\usepackage{graphicx}
\usepackage{dcolumn}
\usepackage{bm}
\usepackage{amsmath, bbm}

\usepackage[hypertex]{hyperref}
\newcommand{\nc}{\newcommand}
\nc{\be}{\begin{eqnarray}}
\nc{\ee}{\end{eqnarray}}
\nc{\bea}{\begin{eqnarray}}
\nc{\eea}{\end{eqnarray}}
\nc{\bean}{\begin{eqnarray*}}
\nc{\eean}{\end{eqnarray*}}
\nc{\mb}{\mbox}
\nc{\rnc}{\renewcommand}
\nc{\vk}{\mb{\bf k}}
\nc{\vp}{\mb{\boldmath$p$}}
\nc{\rr}{\mb{\boldmath$r$}}
\nc{\vR}{\mb{\boldmath$R$}}
\nc{\vz}{\hat {\mb{\bf z}}}
\nc{\vj}{\mb{\boldmath$j$}}
\nc{\vg}{\mb{\boldmath$g$}}
\nc{\vE}{\mb{\boldmath$E$}}
\nc{\vB}{\mb{\boldmath$B$}}
\nc{\vH}{\mb{\boldmath$H$}}
\nc{\vM}{\mb{\boldmath$M$}}
\nc{\vP}{\mb{\boldmath$P$}}
\nc{\vS}{\mb{\boldmath$S$}}
\nc{\x}{\mb{\boldmath$x$}}
\nc{\A}{\mb{\boldmath$A$}}
\nc{\va}{\mb{\boldmath$a$}}
\nc{\vq}{\mb{\boldmath$q$}}
\nc{\vn}{\mb{\boldmath$n$}}
\nc{\vs}{\mb{\boldmath$\sigma$}}
\nc{\vt}{\mb{\boldmath$\tau$}}
\nc{\vpi}{\mb{\boldmath$\pi$}}
\nc{\nab}{\bm{\nabla}}
\nc{\X}{\sf x}

\begin{document}

\title{Origin of giant bulk Rashba splitting: Application to BiTeI}
\author{M. S. Bahramy$^1$}
\email{bahramy@riken.jp}
\author{ R. Arita$^{1,2}$} 
\author{N. Nagaosa$^{1,2}$}
\affiliation{$^1$Correlated Electron Research Group (CERG), RIKEN Advanced Science Institute, Wako, Saitama 351-0198, Japan \\
$^2$Department of Applied Physics, University of Tokyo, Tokyo 113-8656, Japan}
\date{\today}

\begin{abstract}
We theoretically propose the necessary conditions for realization of giant Rashba splitting in bulk systems. In addition to (i) the large atomic spin-orbit interaction in an inversion-asymmetric system, the following two conditions are further  required; (ii) a narrow band gap, and (iii) the presence of top valence and bottom conduction bands of symmetrically the same character.  As a representative example, using the first principles calculations, the recently discovered giant bulk Rashba splitting system BiTeI is shown to fully fulfill  all these three conditions. Of particular importance, by predicting the correct crystal structure of BiTeI, different from what has been believed thus far,  the third criterion is demonstrated to be met  by  a {\it negative} crystal field splitting of the top valence bands. 

\end{abstract}

\maketitle

Exploiting the spin degree of freedom of the electrons  is one of the primary goals in the rapidly growing field of  spintronics. A promising candidate to achieve this goal is so-called Rashba effect,  which relies on the  spin-orbit interaction (SOI) of the carriers  in an inversion ($I$) asymmetric environment~\cite{rashba}. This effect has been experimentally observed for a number of non-magnetic metallic surfaces ~\cite{lashell,ast, koroteev} and also demonstrated to exist at the interface of the semiconductor hetrostructures~\cite{nitta}.  
For most of these systems, the level of Rashba spin  splitting (RSS) are found to be rather small  (at most several meV). However, there have been a few exceptions, e.g. the  Bi-covered Ag(111) surface, for which the angle resolved  photoemisson spectroscopy (ARPES) has revealed a  giant RSS of the order of 200 meV~\cite{ast}. Following these discoveries, attempts have been made to realize RSS in three-dimensional systems, as such systems are expected to be an ideal laboratory for exploring many novel phenomena, e.g.  the spin Hall effect and resonance enhanced magneto-optical conductivity.   

BiTeI, a polar layered semiconductor, has been very recently revisited from this point of view. The ARPES measurements clearly show a gigantic RSS among the lowest conduction bands (LCB's) in the bulk BiTeI~\cite{Ishizaka}, leading to a substantial shift, $k_{CBM}=\pm 0.05$ \AA$^{-1}$, in the position of conduction band minimums (CBM's). The level of spin splitting obtained at CBM astonishingly reach to  $\sim 0.4$ eV lying among the highest discovered so far. The corresponding  Rashba energy $E_R=E_{CBM}-E_{0}$, where $E_0$ indicates the energy of the two LCB's at their crossing point, is found to be over 100 meV. The spin-resolved ARPES  measurements further reveals that these bands are fully spin-polarized, 
reassuring the entire system is indeed subject to a giant RSS~\cite{Ishizaka}.

Motivated by this discovery, we have performed a theoretical study based on the perturbative ${\bf k}\cdot{\bf p}$ formalism and backed by the  first-principles calculations and  group theoretic analysis to investigate  the origin of giant RSS in bulk materials. Through this study, we have identified three conditions required for realization of this intriguing phenomenon. These conditions are shown to be closely related to the relative ordering and symmetry character of the bands near Fermi level,  $E_F$. As a representative case, BiTeI is shown to fully meet all these conditions, owing to its unusual electronic band structure near   $E_F$.

 As a starting point, we describe the general ${\bf k}\cdot{\bf p}$ Hamiltonian via perturbation theory (PT). Given the solution $H({\bf k_0})$ at ${\bf k}={\bf k_0}$, it can be expressed for nearby ${\bf k}$ as,
\begin{equation}
\label{hamil}
H({\bf k})=H({\bf k_0})+\frac{\hbar^2q^2}{2m_0}+\frac{\hbar}{m_0}{\bf q}\cdot{\bf p}+H^{(1)}+H^{(2)}
\end{equation}
\begin{equation}
\label{h1}
H^{(1)}=\frac{\hbar^2}{4m_0^2c^2}(\nabla V\times{\bf q})\cdot{\boldsymbol \sigma},~~H^{(2)}=\frac{\hbar}{4m_0^2c^2}(\nabla V\times{\bf p})\cdot{\boldsymbol \sigma}.
\end{equation}
Here, $V$, ${\boldsymbol \sigma}$ and {\bf p} denote the crystal potential, Pauli matrices and the momentum operator, respectively, and ${\bf q}={\bf k}-{\bf k_0}$. Considering only the linear-in-$k$ spin splittings, one can show that they can arise due to $H^{(1)}$ by the use of the first-order PT or the coupling between perturbing terms $\frac{\hbar}{m_0}{\bf q}\cdot{\bf p}$ and $H^{(2)}$ in the second order PT. The spin splitting arising from $H^{(1)}$ is, however, expected to be much less than that coming from $H^{(2)}$, and hence, unlikely to cause any giant RSS (i.e. $> 100$ meV). This is because $H^{(1)}$ ($H^{(2)}$), as reflected by its ${\bf k}$-dependence (${\bf p}$-dependence),  originates  from the crystal (atomic orbital) momentum. Since the velocity of the electron in its atomic orbit is far greater than the velocity of a wave packet, the spin splitting is accordingly expected to be strongly dominated by $H^{(2)}$~\cite{kane}.  
The respective second order perturbative correction in energy is given by~\cite{voon,voon-book},
\begin{equation}
\label{hpert}
\Delta \varepsilon_{m}^{(2)}({\bf k})=\frac{\hbar}{m_0} \sum_{{n} \ne m} \frac {\langle u_m \arrowvert H^{(2)} \arrowvert u_{{n}} \rangle\langle u_{{n}} \arrowvert {\bf q}\cdot{\bf p} \arrowvert u_m\rangle +c.c.}{\varepsilon_m-\varepsilon_{{n}}}
\end{equation}
where $u_i$ and $\varepsilon_i$ are the eigenstate and eigenenergy corresponding to the state $i$ at ${\bf k_0}$, respectively, and  $c.c.$ stand for the complex conjugation. 

Equation~(\ref{hpert}) clearly indicates that the level of spin splitting is directly dependent on three conditions:  (i) the strength of the  spin orbit interaction (represented by $H^{(2)}$) (ii) the energy difference between the neighboring states  $m$ and $n$, and (iii) the symmetry character of their corresponding eigenstates, determining if $\langle u_m \arrowvert H^{(2)} \arrowvert u_{{n}} \rangle$ is symmetrically allowed or not. In brief,  states energetically close to each other and symmetrically of the same character, can effectively couple with each other, and hence produce a large spin splitting, of course  if the host atoms maintain a strong SOI. The first condition is of course a rather obvious requirement, already well-known in the context of surface RSS~\cite{oguchi}. The other two are, however, less trivial and require more attention. In the case of semiconducting bulk materials, they can be satisfied if the band gap is sufficiently narrow and, more importantly, if both  LCB's and the highest valence bands (HVB's) are symmetrically the same.  The last condition  requires an anomalous ordering of the bands near $E_F$ which is not usually allowed in conventional semiconductors. However,  some polar semiconductors can exceptionally meet this criterion due to the $negative$ crystal field splitting (CFS) of their top valence bands (TVB's). BiTeI is one such example, which in the following will be shown to fulfill all these three  conditions and thereby exhibiting a giant bulk RSS.

\begin{figure}[pt]
\begin{center}
\rotatebox{0}{\includegraphics[width=3 in]{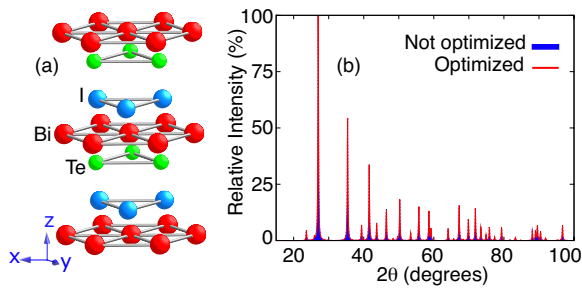}}
\end{center}
\caption{\label{fig:struc} (Color online) (a) Crystal structure of BiTeI and (b) simulated XRD pattern for the optimized and non-optimized structures.}
\end{figure} 

\begin{figure*}[pt]
\begin{center}
\rotatebox{0}{\includegraphics[height=2 in]{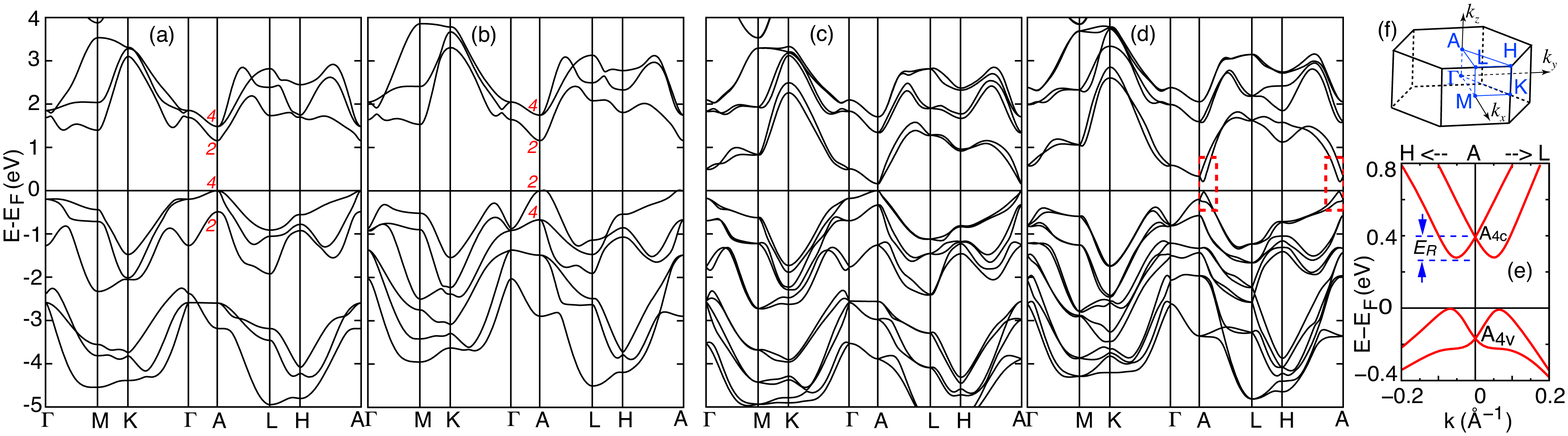}}
\end{center}
\caption{\label{fig:xpband}  (Color online) Calculated electronic band structures for the non-optimized structure (a) without SOI and (c) with SOI and for the optimized structure (b) without SOI and (d)  with SOI. (e) An scaled-up view of band dispersions along $H$-$A$-$L$ direction for the optimized structure. (f) Corresponding high symmetry $k$-points in the hexagonal Brillouin zone of BiTeI.  In (a) and (b), the oblique numbers indicate the band degeneracies at  $A$ point for each structure (see the related discussion). }
\end{figure*}

Having a trigonal structure with the space group $P3m1$ (No. 156), BiTeI is a polar compound in which   Bi, Te and I form stacking layers along the $c$-axis. A characteristic feature of $P3m1$ group, as shown in Fig.~\ref{fig:struc}(a), is the lack of $I$-symmetry. The highest symmetry operation allowed is $C_{3v}$ along the  $c$ axis. The corresponding experimental lattice  parameters $a$ and $c$ are  $4.339$~\AA~and $6.854$~\AA, respectively~\cite{Shevelkov1995}.   
Assuming Bi to be at origin, Te and I have been experimentally proposed to be respectively at  $(2/3, 1/3,  0.6928)$ and $(1/3, 2/3, 0.2510)$ sites~\cite{Shevelkov1995}. However, after a full structural optimization of atomic positions~\cite{wien2k}, we have surprisingly found that Te and I change their positions to $(2/3, 1/3,  0.7482)$ and $(1/3, 2/3, 0.3076)$, respectively. In other words, our predicted Bi-Te (Bi-I) distance is exactly equal with the experimentally proposed Bi-I (Bi-Te) distance. Such a correction, as will be discussed later, leads to a number of fundamental changes in the electronic  structure of BiTeI, including the appearance of a gigantic RSS, similar to what was  observed  by ARPES~\cite{Ishizaka}. 
The failure of XRD experiment might be due to the fact that Te and I both have nearly the same ionic radii (133 pm and 131 pm, respectively ) and atomic charges (52 and 53, respectively). Accordingly they likely produce rather indistinguishable features in the XRD pattern. Indeed  the simulated  XRD patterns ~\cite{crystaldifract}, as shown in Fig.~\ref{fig:struc}(b), turn out to be nearly identical for the both structures. Thus, the utilization of  more sophisticated experimental techniques seems to be necessary for the proper identification of atomic positions in BiTeI.
In the rest of this letter, we compare in detail the electronic structures of both the non-optimized and optimized  BiTeI , to asses  if the above mentioned conditions are indeed required for a giant bulk RSS. 

Figure~\ref{fig:xpband}  shows the respective band structures for the both systems.  In the absence of SOI, they  show a semiconducting behavior  with an energy gap $E_G\sim1.2$ eV. As shown in Fig.~\ref{fig:xpband}(a) and (b), the lowest $E_G$ is commonly found to be not at the BZ center, but at $A$ point where $k_x=k_y=0$ and $k_z=\pi$ (see Fig.~\ref{fig:xpband}-(f)). Up to 4 eV above  $E_F$, the (six) conduction bands are predominated by Bi-$6p$ states, whereas the Te-$5p$ and I-$5p$ most strongly contribute to the (twelve) valence bands down to -5 eV below $E_F$.
Another important similarity  is the presence of a rather large CFS among these  bands. Without SOI, for any $k$-point other than those along $\Gamma$-$A$  or the ones with accidental symmetry, all the $p$-type bands split into doubly degenerate bands. Along $\Gamma$-$A$, such states are allowed to form either 2-fold or 4-fold degenerate bands. Such a trend of CFS  can be well described using the group theory.  
Within $P3m1$ group, at the BZ center and along $\Gamma$-A all the $k$-points have  $C_{3v}$ symmetry.  Without spin, all the bands at $A$ are thus transformed according to one of the single group representations of $C_{3v}$, (a complete character table for $C_{3v}$ can be found in the supplemental file). Of our particular interest are $\{s, p_z\} \rightarrow A_1$ and $ \{p_x,p_y\} \rightarrow A_3$,
explaining why the $p$-type valence and conduction bands along $\Gamma$-$A$ are either two-fold or four-fold degenerate. For the other internal BZ points
all the single representations are one dimensional and hence non-degenerate except for spin.

Comparing the ordering of CFS of conduction bands at $A$, one can notice a similar trend, that is, $A_1$-$A_3$ in the increasing order of energy. The same trend holds for TVB's in the non-optimized structure. However, in the optimized structure the ordering of TVB's is opposite, i.e. $A_3$-$A_1$  (see Fig.~\ref{fig:xpband}(b)). This accordingly implies the existence of a {\it negative} CFS among this group of bands. As a result, both the LCB's and HVB's turn out to be $A_1$ ($p_z$) type and 2-fold degenerate. It is to be noted that such a negative CFS has already been observed for a number of chalcopyrite-type semiconductors, e.g. CdSnP$_2$~\cite{shay} and CuAlS$_2$~\cite{Jayalakshmi} as well as AlN~\cite{taniyasu}. The reason has been attributed to the strong ionicity of the atomic bondings, leading to a substantial  structural distortion along their high symmetry axis.

Turning on SOI, the band structures undergo yet another drastic change.  As shown in Fig.~\ref{fig:xpband}(c) and (d),   $E_G$ is closed down to $\sim 0.28$ eV, nearly 5 times smaller than that obtained without SOI. This is mainly due to the strong SOI of Bi which shifts downward the $j=1/2$ bands by nearly $-2\Delta_{so}$, where $\Delta_{so}$ denotes the atomic SOI energy (for Bi it is found to be $\sim0.5$ eV).  Despite this similarity, the trend of spin splitting near $E_F$ is strikingly different between the two system. A comparison between Fig.~\ref{fig:xpband}(c) and (d) clearly reveals that in the optimized structure a huge RSS takes place among both LCB's and HVB's at $A$ point, whereas the non-optimized structure fails to yield such a feature. As shown in Fig.~\ref{fig:xpband}(e), in the optimized structure,  CBM and VBM are both  shifted by  nearly $k_{CBM}=k_{VBM}=\pm 0.05$ \AA$^{-1}$ from $A$ in the $(k_x,k_y)$ plane with an  $E_R=113$ meV, in excellent agreement with the ARPES data~\cite{Ishizaka}. Such a good agreement is a strong indication that our predicted structure is indeed correct.

\begin{figure}[pb]
\begin{center}
\rotatebox{0}{\includegraphics[width=3 in]{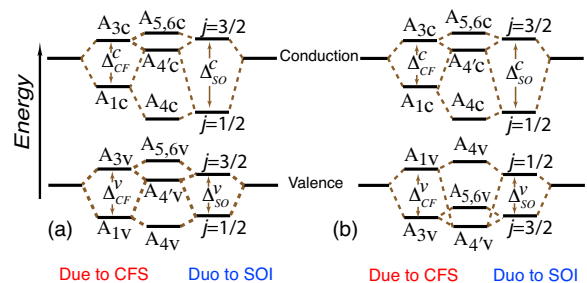}}
\end{center}
\caption{\label{fig:bdiag} (Color online) Diagrammatic representation of band splitting due to the crystal field splitting (CFS)  and spin-orbit interaction (SOI) and their combination in (a) non-optimized and (b) optimized BiTeI.}
\end{figure}

 Having confirmed the existence of a giant RSS in bulk BiTeI,  we next address the main question: whether this effect arises due to the fulfillment of all the three conditions, pointed out earlier. As already might be understood, the both optimized and non-optimized structures meet the first two criteria, namely the strong SOI in an $I$-asymmetric environment and narrow band gap. However, the last condition is only satisfied in the optimized structure as its both LCB's and HVB's are symmetrically the same. 
To be more specific, upon introduction of SOI, the previously defined single group representations $A_1$ and $A_3$ transform to their double group counterparts (see the character table in supplemental file).  In $C_{3v}$ space  the transformation is such that $A_1 \rightarrow  A_4$ and $A_3 \rightarrow A_4\oplus A_5\oplus A_6$.
In other words, $A_1$ transforms to $A_4$, whereas $A_3$ is split into two 2-fold bands $A_4$ and $A_5\oplus A_6$ ( the latter is  hereafter simplified as $A_{5,6}$). Since TVB's of optimized structure undergo a negative CFS, the corresponding HVB's and LCB's are  both of  $A_4$ character. Fig.~\ref{fig:bdiag} schematically shows the effects of CFS and SOI. 

   Turning back to Eqs.~(\ref{hamil}) and (\ref{h1}), one can use the method of invariants~\cite{voon-book} to find an effective spin-splitting Hamiltonian linear in terms of $k$. For $C_{3v}$ symmetry, it is easy to show that, the only possible choice is $H_q\propto (\sigma_x q_y-\sigma_y q_x)$
because both $(q_x,q_y)$ and $(\sigma_x,\sigma_y)$ belong to the $A_3$ representation. $H_q$ is clearly  a Rashba-type Hamiltonian, giving the following spin-split energies $\Delta \varepsilon_m=\pm\alpha_m\sqrt{q_x^2+q_y^2}$. As described in the beginning, $\Delta \varepsilon_m$ is expected to be dominated by $\Delta \varepsilon_{m}^{(2)}$.  
We can thus pursue with PT to qualitatively determine   the second order correction in Rashba parameter $\alpha_m$, denoted as $\alpha_{m}^{(2)}$. 
With the help of group theory, it turns out that in Eq.~(\ref{hpert}) $\langle u_m \arrowvert H^{(2)} \arrowvert u_{{n}} \rangle\ne 0$,  if and only if both $u_m$ and $u_n$ are of $A_4$ character, implying that linear-in-k splitting is symmetrically forbidden for  $A_{5,6}$ bands. Accordingly, we just need to consider  ${{u_i}}=A_4$ and $A_{4'}$ (see Fig.~\ref{fig:bdiag}).
To derive an analytical form for $\alpha_{m}^{(2)}$ from Eq.~(\ref{hpert}), analogous to Gutche and Jane basis~\cite{gutsche,voon,voon-book}, we define  
$\arrowvert A_4\uparrow\downarrow\rangle=\pm \sqrt{\frac{1-q_4^2}{2}}\arrowvert (x \pm iy)\downarrow\uparrow\rangle+q_4\arrowvert Z\uparrow\downarrow\rangle$, and $\arrowvert A_{4'}\uparrow\downarrow\rangle=\pm \sqrt{\frac{q_4^2}{2}}\arrowvert (x \pm iy)\downarrow\uparrow\rangle-\sqrt{1-q_4^2}\arrowvert Z\uparrow\downarrow\rangle$
, where $q_4$ is  to enforce the orthogonality of the basis sets and $+~ (-)$ sign corresponds to spin $\uparrow$ (spin  $\downarrow$). The $A_1$-like basis $\arrowvert Z\rangle$ is defined as  $\arrowvert Z\rangle=q_s\arrowvert s\rangle+q_z\arrowvert z\rangle$ with $q_s^2+q_z^2=1$.  

To simplify our derivations, we recall   the fact that the states with small $({\varepsilon_m-\varepsilon_{{n}}})$ dominate   $\Delta \varepsilon_{m}^{(2)}({\bf k})$ and, thus $\alpha_{m}^{(2)}$. From Fig.~\ref{fig:xpband}(c) and (d), it is evident that $\arrowvert \varepsilon_{4c}- \varepsilon_{4'c} \arrowvert>1.30 $ eV and $\arrowvert  \varepsilon_{4v}- \varepsilon_{4'v} \arrowvert>1.15 $ eV. On the other hand, for the optimized and non-optimized structures, the respective $\Delta  \varepsilon_{4,4}=\varepsilon_{4c}- \varepsilon_{4v}$ and $\Delta  \varepsilon_{4,4'}= \varepsilon_{4c}-\varepsilon_{4'v}$ are both below 0.5 eV and very close to their $E_G$. Thus, as a good approximation for the former (latter),  $\alpha_{4c}^{(2)}=\alpha^{(2)}_{4c,4v}$ ($\alpha_{4c}^{(2)}=\alpha^{(2)}_{4c,4'v}$) with,
\begin{eqnarray}
\label{a2}
\alpha^{(2)}_{4c,4v}&=&\frac{\Delta_{4,4}}{\sqrt{2}\Delta \varepsilon_{4,4}}[q_{4v}\sqrt{1-q^2_{4c}}P_{x,Z}+q_{4c}\sqrt{1-q^2_{4v}}P_{Z, x}]  \nonumber \\
\alpha^{(2)}_{4c,4'v}&=&\frac{\Delta_{4,4'}}{\sqrt{2}\Delta \varepsilon_{4,4'}}[\sqrt{(1-q^2_{4v})(1-q^2_{4c})}P_{x, Z}-q_{4c}q_{4v}P_{Z, x}] 
\end{eqnarray}
where, $\Delta_{i,j}\equiv \langle A_{ic}\uparrow\arrowvert  H^{(2)} \arrowvert A_{jv}\uparrow\rangle$ and $P_{x, Z}=-i\hbar\langle x_c\arrowvert \partial /\partial x \arrowvert Z_v \rangle$.

Constructing a set of  maximally localized Wannier functions~\cite{souza,mostofi,kunes} for the  conduction and valence bands around $E_F$, we have estimated $q^2_{4v}$ and $q^2_{4c}$ for the optimized and non-optimized structures.  For the latter, $q^2_{4v}=0.55$ and $q^2_{4c}=0.43$, both very close to the critical value 0.5. Assuming $P_{x,Z}=P_{Z, x}$, one can immediately find from Eq.~(\ref{a2}) that $\alpha_{4c}^{(2)}\simeq 0$. Since $\alpha^{(2)}_{4'v,4c}=-\alpha^{(2)}_{4c,4'v}$, then $\alpha_{4'v}$ is also expected to be nearly zero. This clearly explains why the non-optimized structure shows almost no spin splitting among its HVB's and LCB's near the $A$ point. As for the optimized structure, the situation is completely different. Here, $q^2_{4v}=0.886$ and $q^2_{4c}=0.5$, implying that the HVB's are predominantly $Z$-type. Consequently,  $\alpha_{4c}^{(2)}$ and $\alpha_{4v}^{(2)}$ turn out to have appreciable values with equal magnitudes but opposite signs. In other words under this situation, for both HVB's and LCB's, the absolute value of $\alpha$ can be nearly the same, but their signs are always opposite. That's exactly what we can see in the optimized BiTeI as it also shows similar trend of spin-splitting among its LCB's and HVB's such that their corresponding $k_{CBM}$ and $k_{VBM}$ are almost at the same place. Here, It is important to emphasize  that such a second order perturbative RSS is a direct result of an (i) anomalous ordering of top valence bands due to the existence of negative CFS which allows the adjacent $A_{4v}$ and $A_{4c}$ to be symmetrically of the same character and, hence, to be coupled with each other thorough a perturbative Rashba-like Hamiltonian. Due to (ii) the  large SOI of Bi leading to  (iii) substantial band gap narrowing , such a coupling can be effectively very strong. These are the three key factors for realization of  giant RSS in BiTeI, and very likely any other giant bulk Rashba splitting material. For the other candidates a negative CFS  can either intrinsically exist due to the strong anisotropic ionicity of their atomic bondings or  be externally produced e.g. through a pressure-induced structural distortion.  

In summary, in this study we combined the first-principles calculations with a group theoretic analysis to investigate the origin of giant Rashba splitting in bulk systems. As a representative case, It was shown that in BiTeI the interplay between the giant SOI of Bi and effectively large negative CFS of the TVB's led to a substantially strong coupling between the narrowly separated HVB's and LCB's via a perturbative Rashba-like Hamiltonian. Such conditions were expected to be vital for realization of giant RSS in other bulk candidates, also.  
 
This study was supported by Funding Program for World-Leading Innovative R\&D on Science and Technology (FIRST program) on "Quantum Science on Strong Correlation".

\end{document}